\documentclass[a4paper,11pt]{article}
\pdfoutput=1 

\usepackage{jcappub} 

\usepackage[T1]{fontenc} 
\usepackage[utf8]{inputenc}

\title{The redshift distribution of cosmological samples: a forward modeling approach}

\author{J\"org Herbel,}
\author{Tomasz Kacprzak,}
\author{Adam Amara,}
\author{Alexandre Refregier,}
\author{Claudio Bruderer}
\author{and Andrina Nicola}

\affiliation{Institute for Astronomy, Department of Physics, ETH Z\"urich, Wolfgang-Pauli-Strasse 27, 8093 Z\"urich, Switzerland}

\emailAdd{joerg.herbel@phys.ethz.ch}

\abstract{Determining the redshift distribution $n(z)$ of galaxy samples is essential for several cosmological probes including weak lensing. For imaging surveys, this is usually done using photometric redshifts estimated on an object-by-object basis. We present a new approach for directly measuring the global $n(z)$ of cosmological galaxy samples, including uncertainties, using forward modeling. Our method relies on image simulations produced using \textsc{UFig} (Ultra Fast Image Generator) and on ABC (Approximate Bayesian Computation) within the \textit{MCCL} (Monte-Carlo Control Loops) framework. The galaxy population is modeled using parametric forms for the luminosity functions, spectral energy distributions, sizes and radial profiles of both blue and red galaxies. We apply exactly the same analysis to the real data and to the simulated images, which also include instrumental and observational effects. By adjusting the parameters of the simulations, we derive a set of acceptable models that are statistically consistent with the data. We then apply the same cuts to the simulations that were used to construct the target galaxy sample in the real data. The redshifts of the galaxies in the resulting simulated samples yield a set of $n(z)$ distributions for the acceptable models. We demonstrate the method by determining $n(z)$ for a cosmic shear like galaxy sample from the 4-band Subaru Suprime-Cam data in the COSMOS field. We also complement this imaging data with a spectroscopic calibration sample from the VVDS survey. We compare our resulting posterior $n(z)$ distributions to the one derived from photometric redshifts estimated using 36 photometric bands in COSMOS and find good agreement. This offers good prospects for applying our approach to current and future large imaging surveys.}

\begin{document}
\maketitle
\flushbottom

\section{Introduction}

The $\Lambda$CDM model, which has become our standard model in cosmology, successfully captures the physical evolution of the Universe. Yet, the dark components of this model, i.e. dark matter and dark energy, which together account for the majority of the energy density of the Universe, are still poorly understood (e.g. \cite{Spergel2015} and references therein). Therefore, gaining insight into the dark sector of the Universe is one of the major goals of modern cosmology.

In the era of precision cosmology, we are now able to measure the properties of dark energy and dark matter using several cosmological probes. Examples include weak gravitational lensing (cosmic shear), galaxy clustering, baryonic acoustic oscillations, type Ia supernovae and the cosmic microwave background (CMB) (e.g. \cite{Albrecht2006}). All of these probes can be used to make cosmological measurements on their own, and yield stronger constraints when combined. However, they are affected by systematic errors that need to be controlled carefully in order to obtain reliable results (see e.g. \cite{Amara2008} for systematics concerning cosmic shear and \cite{Prada2016} for baryonic acoustic oscillations).

For instance, several of these probes, such as cosmic shear, projected galaxy clustering and CMB cross-correlations rely on a precise determination of the redshift distribution, $n(z)$, of a target galaxy sample (e.g. \cite{Nicola2017} and references therein). The target sample is typically constructed from a series of cuts that are optimized for the specific probe. A careful quantification of the uncertainty on $n(z)$ of the sample is also required along with its propagation through to the cosmological parameter constraints \cite{Amara2007,Abdalla2008,Bordoloi2010,Bordoloi2012}. However, obtaining $n(z)$ for ongoing and upcoming large photometric galaxy surveys such as the Dark Energy Survey\footnote{\url{http://www.darkenergysurvey.org}} (DES), the Kilo-Degree Survey\footnote{\url{http://kids.strw.leidenuniv.nl}} (KiDS) and the survey of the Large Synoptic Survey Telescope\footnote{\url{https://www.lsst.org}} (LSST) is a major challenge. Due to the area and the depth covered by these experiments, any corresponding spectroscopy would be too cost- and time-intensive. Therefore, one has to rely on redshift information derived from photometric flux measurements using a limited number of broad band filters (see e.g. \cite{Bonnett2016}).

One approach to photometric redshift estimation is based on template fitting (e.g. \cite{Benitez2000, Brammer2008}), which uses a library of galaxy spectra to fit the observed multi-band fluxes, thereby determining redshifts. Another approach has been to use machine learning methods such as neural networks and boosted decision trees (e.g. \cite{CarrascoKind2013, Sadeh2016}). This latter method relies on using a representative sample with known redshifts to train the machine learning algorithms.

Both template fitting and machine learning methods first estimate the redshifts of individual objects, which can then be combined to produce estimates of $n(z)$ of the target sample. However, given that the cosmological probes mentioned above specifically require the global $n(z)$ of the sample, we propose here a method that bypasses object-by-object estimates and directly determines the population properties. Our method therefore aims to obtain a full posterior for $n(z)$, i.e., a family of curves representing the possible redshift distributions. 

The method is based on a forward modeling approach that we apply in a Bayesian framework, using Approximate Bayesian Computation (ABC). This makes use of wide-field image simulations generated using the Ultra Fast Image Generator (\textsc{UFig}). The approach is an application of the Monte-Carlo Control Loops (\textit{MCCL}, \cite{Refregier2014}) framework. The original description of the \textit{MCCL} scheme focussed on the shear measurement process, which relies on image simulations for calibration purposes, whereas here, we extend the framework to determine redshift distributions.

Using a forward modeling approach allows us to control the astrophysics included in the measurement process. We use a parametric model that relies on a few basic, well founded assumptions. Specifically, the galaxy population is modeled using luminosity functions that evolve with redshift. We use different luminosity functions to describe blue and red galaxies separately. Furthermore, we use parametric models to assign spectra, sizes and radial profiles to galaxies. We also include instrumental and observational effects when generating images. 

The simulated images are analyzed in the same way as the real data. The input parameters are adjusted so that the simulations statistically agree with the data. This yields a family of acceptable models that are consistent with the data. We then apply the same cuts that were used for the real data to the simulations. The redshifts of the galaxies in the resulting simulated samples yield a set of $n(z)$ distributions for the acceptable models.

To demonstrate the method, we use it to determine the redshift distribution of a cosmic shear like galaxy sample from the 4-band Subaru Suprime-Cam data \cite{Capak2007} in the COSMOS field \cite{Scoville2007}. We also complement this imaging data with a spectroscopic calibration sample from the VVDS survey \cite{LeFevre2005, Garilli2008, LeFevre2013}. We compare our resulting posterior $n(z)$ distributions to the one derived from the photometric redshifts estimated by the COSMOS collaboration using 36 photometric bands.

The paper is structured as follows: Section \ref{sec:method} presents the general method used to measure $n(z)$ and section \ref{sec:model} describes our modeling scheme. In section \ref{sec:test-case}, we introduce the test case that is used to demonstrate our method, while section \ref{sec:adjusting-ufig} describes the details of the implementation. Our results are presented in section \ref{sec:results}. Our conclusions are summarized in section \ref{sec:conclusions}. Throughout this work, we use a standard $\Lambda$CDM cosmology with $\Omega_m = 0.3$, $\Omega_\Lambda = 0.7$ and $H_0 = 70 \, \text{km} \, \text{s}^{-1} \, \text{Mpc}^{-1}$.

\section{Method}
\label{sec:method}

In this section, we describe the details of our method for measuring the $n(z)$ distribution of a target sample of galaxies derived from imaging data, where the target sample is defined by a series of cuts. Our method relies on a forward modeling approach. Specifically, we statistically compare simulated images, generated using \textsc{UFig}, with the real data to find a model space for which our simulations are in good agreement with the data. By applying exactly the same analysis process and cuts to the simulations, as was done for data, we produce a series of simulated target samples. By cross-matching the detected objects in the simulations with the \textsc{UFig} input catalogs that include redshifts, we can produce an $n(z)$ curve for each of the simulated target samples. This family of curves gives a posterior distribution for the $n(z)$ of the real target sample.

The model used in the simulations includes parameters that describe the galaxy population as a function of redshift, instrumental features and observational effects. Our goal is to infer the values of these parameters in probabilistic Bayesian way (see e.g. \cite{VonToussaint2011} for a review). However, the challenge is that the likelihood for this problem is not available. That is to say that when comparing two catalog samples where each object has many properties, there is no clear empirical likelihood that can be calculated. The same holds for image level comparisons.

We avoid this problem by using Approximate Bayesian Computation (ABC, \cite{Robert2016, Akeret2015}).  ABC allows us to perform a Bayesian analysis in the case where the likelihood is not tractable. The method works by stochastically sampling from the prior model parameter space to create a suite of simulated data. These simulations are then compared against survey data and the difference is calculated using distance metrics. The Bayesian posterior distribution is approximated by only accepting samples for which the distance metrics are smaller than some thresholds. The generation of the large number of simulations needed to do this is made possible by the computational speed of \textsc{UFig}.

The fact that ABC yields an approximate posterior distribution allows us to quantify the uncertainty on the $n(z)$ distribution. The goodness of the approximation to the true posterior generally depends on the distance metrics and the chosen thresholds. By lowering the thresholds, the accuracy of the approximation improves. However, this increases the computational costs, as more model evaluations are required to find a sufficient number of points that fulfill the acceptance criterion. In practice, one has to find a balance between precision and computational costs.

The method that we present in this paper is an implementation of the Monte-Carlo Control Loops (\textit{MCCL}, \cite{Refregier2014}) framework. The \textit{MCCL} scheme originally focussed on shear measurements and provided a way to calibrate these measurements and rigorously handle systematic errors using forward modeling. A first implementation was presented in \cite{Bruderer2016}. We extended the \textit{MCCL} framework to measure $n(z)$. As in the original description, there are three iterative loops to
\begin{enumerate}
\item tune the image simulations to statistically match the data on which the measurement is performed, 
\item calibrate the measurement using the adjusted image simulations,
\item explore the effects of systematics via Monte-Carlo methods.
\end{enumerate}

In this paper, we apply the \textit{MCCL} scheme to wide-field imaging data to measure $n(z)$ including uncertainties. Control loop 2 is trivial in our case, since the measurement step consists of using the redshifts assigned to galaxies by the image simulation software to obtain $n(z)$. Furthermore, we effectively perform control loops 1 and 3 within one step, since we do not simply find a best-fit point, but also characterize the uncertainty on the former, which yields a posterior distribution of the possible $n(z)$ curves.

\section{Model}
\label{sec:model}

\subsection{Image simulations}

The original implementation of \textsc{UFig} described in \cite{Berge2013} was designed to render \textit{r}-band images. We extended its functionality to generate images in arbitrary filter bands, including redshifts and colors for galaxies. The basic operating principle is to first generate randomly drawn catalogs of sources using models of the intrinsic populations of galaxies and stars and to then render the pixelated light profiles. \textsc{UFig} operates on the level of individual image tiles and can simulate full surveys image-by-image. We also include observational and instrumental effects such as pixel noise, the PSF and saturation in the rendering process. In the following, we will describe the newly added features.

\paragraph*{Galaxies}

\begin{figure}[tbp]
\centering
\includegraphics[width=\textwidth]{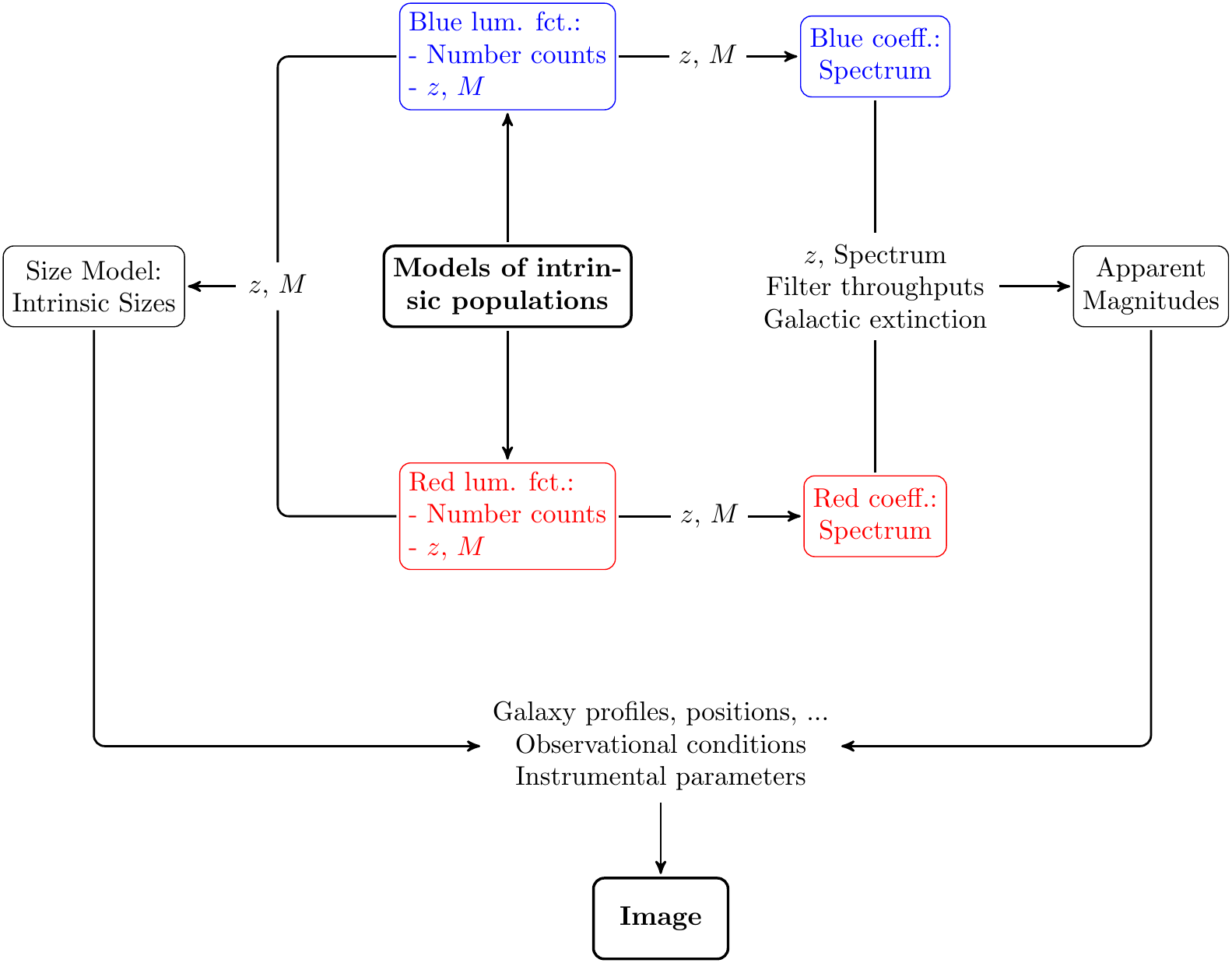}
\caption{Schematic illustration of the process of sampling galaxies in \textsc{UFig}. The figure illustrates our usage of intrinsic models of the galaxy population to render wide-field images.}
\label{fig:model-galaxies}
\end{figure}

The process of drawing galaxies is illustrated in figure \ref{fig:model-galaxies}. To sample the population of the former, we rely on luminosity functions. They are observationally well determined quantities which measure the comoving space density of galaxies, see \cite{Johnston2012} for a review and \cite{Beare2015} for a compilation of recent results. We discriminate between blue and red galaxies and use separate luminosity functions to model these two populations. This split into blue, star-forming and red, passive galaxies is very common in the literature on galaxy populations and a large number of results for the luminosity functions describing the two populations is available. In appendix \ref{sec:nyu-vagc}, we describe the split of SDSS galaxies into blue and red galaxies. We use specific star formation rates to distinguish the two populations.

Furthermore, we evolve the luminosity functions with redshift to capture the evolution of galaxies. To generate a galaxy catalog, we first compute the expected number counts within a given volume and magnitude range (separately for blue and red objects) and then sample redshifts and absolute magnitudes according to the corresponding luminosity function (either the blue or the red one).

To compute apparent magnitudes in arbitrary filter bands, we assign a spectrum to each galaxy using a parametric model where we again allow for evolution with redshift. This enables us to assign apparent magnitudes to each source, whereby we include reddening due to interstellar dust. Furthermore, we draw the intrinsic physical sizes of galaxies based on their absolute magnitude and transform to apparent sizes using the distance to each galaxy.

To finally render galaxies on the simulated image, we assign them positions and light profiles, whereby we retained the prescriptions explained in \cite{Berge2013}. Taking into account observational conditions such as the PSF and instrumental parameters such as the pixel scale and the gain, we evaluate each light profile on the pixelated image to obtain realistic wide-field data.

\paragraph*{Stars}

Stars are sampled using the Besançon model of the Milky Way. Its features and the implementation are described in \cite{Robin2003}. This model predicts number counts and magnitudes. The light profiles of stars are given by the PSF and can be adjusted according to the observational conditions. We again refer to \cite{Berge2013} for details, also concerning the assignment of positions. Respecting the parameters of the corresponding instrument, the stellar profiles are eventually evaluated on the image.

Since we aim at constraining $n(z)$, it is not strictly necessary to include stars in the simulations. However, simulating the stars mimics, for example, source confusion effects that are present in the data, which makes the simulations more realistic. Furthermore, we use the simulated stars to verify that the PSF sizes we use in the simulations are consistent with the data.

\subsection{Galaxy population}

\subsubsection{Counts, redshifts and absolute magnitudes}
\label{subsubsec:gal-counts-z-abs-mag}

The number of galaxies rendered in \textsc{UFig} is computed from the galaxy luminosity function $\phi$, which is defined as the number of galaxies $N$ per comoving volume $V$ and absolute magnitude $M$:

\begin{equation}
\phi(z, M) = \frac{dN}{dM \, dV},
\end{equation}

where $z$ denotes redshift. The cosmological principle ensures that the luminosity function is independent of the direction in the sky when averaging over sufficiently large areas. The functional form of $\phi$ is taken to be a Schechter function, 

\begin{equation}
\phi = \frac{2}{5} \log(10) \phi_*  \, 10^{\frac{2}{5}(M_* - M) (\alpha + 1)} \, \exp\left[ -10^{\frac{2}{5}(M_* - M)} \right],
\end{equation}

where $\phi_*$, $M_*$ and $\alpha$ are free parameters that need to be adjusted using observational data. The blue and red populations of galaxies are described by different values of these three parameters. To evolve $\phi$ with redshift, we fix $\alpha$ (to different values for blue and red galaxies) and parameterize the evolution of $M_*$ and $\phi_*$ according to

\begin{align}
M_*(z) &= a_M \, z + b_M, \label{eq:lum-fct-redshift-evolution-m*} \\ 
\phi_*(z) &= b_\phi \, \exp \left( a_\phi \, z \right), \label{eq:lum-fct-redshift-evolution-phi*}
\end{align}

where $a_M$ and $b_M$ are the slope and the intercept that set the evolution of $M_*$ and $a_\phi$ and $b_\phi$ are the exponential decay rate and the amplitude that determine the evolution of $\phi_*$. $M_*$ and $\phi_*$ evolve according to the same functional forms for blue and red galaxies, but the values of $a_M$, $b_M$, $a_\phi$ and $b_\phi$ are different for the two populations. This particular way of evolving $\phi$ is empirically motivated (see section \ref{subsubsec:gal-lum-fct-prior}). To calculate the number of galaxies in a given redshift and magnitude range, we rewrite $\phi$ using the comoving volume element $dV$ given by

\begin{equation}
dV = \frac{d_H d^2_M}{E(z)} \, d\Omega \, dz,
\end{equation}

where $d_H$ is the Hubble distance, $d_M$ the transverse comoving distance, $\Omega$ the solid angle and $E(z) = \left( \Omega_M(1+z)^3 + \Omega_k(1+z)^2 + \Omega_\Lambda \right)^\frac{1}{2}$, see \cite{Hogg1999}. Hence, the number density of galaxies in terms of redshift, absolute magnitude and solid angle can be written as

\begin{equation}
\label{eq:z-mabs-dist}
\varphi = \frac{dN}{dM \, d\Omega \, dz} = \frac{d_H d^2_M}{E(z)} \phi.
\end{equation}
The total number of galaxies in a volume subtended by a solid angle $\Omega$ with redshifts between $z_1$ and $z_2$ and absolute magnitudes between $M_1$ and $M_2$ can be computed from $\varphi$ as
\begin{equation}
N(z_1, z_2, M_1, M_2, \Omega) = \Omega \int\limits_{M_1}^{M_2} \int\limits_{z_1}^{z_2} dM dz \, \varphi. 
\end{equation}

Furthermore, by considering $\varphi$ as an (un-normalized) joint probability distribution for $z$ and $M$, one can assign redshifts and absolute magnitudes to galaxies. We distinguish between blue and red galaxies and use separate luminosity functions to describe their populations.

\subsubsection{Spectral energy distributions}
\label{subsubsec:gal-spectrum-model}

We draw the rest-frame spectral energy distribution $f_\text{e}(\lambda)$ of a galaxy as a linear combination of five template spectra $f_{\text{e}, i}(\lambda)$:

\begin{equation}
\label{eq:spec-flux-density}
f_\text{e}(\lambda) = \sum_i c_i f_{\text{e}, i}(\lambda).
\end{equation}

This particular way of assigning spectra to galaxies is empirically motivated. It has been shown that the photometry of galaxies in the Sloan Digital Sky Survey (SDSS) can generally be well described by the five template spectra that we use. A more detailed description is given in appendix \ref{sec:nyu-vagc}. The corresponding coefficients $c_i$ are sampled from a Dirichlet distribution \cite{Balakrishnan2003} of order five. The reason for choosing this particular model is that after drawing the $c_i$ and calculating $f_\text{e}(\lambda)$, we rescale the spectrum to match the absolute magnitude obtained from the luminosity function. This means that the $c_i$ we assign to a galaxy mainly determine its colors, but not its intrinsic brightness. In fact, only the ratios of the $c_i$ matter, their overall sum is irrelevant. It is therefore sufficient to limit the $c_i$ to a four-dimensional simplex, restricting them to always sum up to 1. This is exactly the support of a Dirichlet distribution of order five, which is the reason why we chose this model.

To allow galaxies to evolve with cosmic time, we furthermore use a redshift-dependent model to assign them spectra. A Dirichlet distribution of order five is parameterized by five parameters $\alpha_i$. These are taken to be dependent on redshift, such that the spectrum we assign to a galaxy depends on where in redshift it is located. Specifically, $\alpha_i$ evolves according to

\begin{equation}
\label{eq:template-coeff-evolution}
\alpha_i(z) = \left( \alpha_{i, 0} \right)^{1 - z/z_1} \, \times \, \left( \alpha_{i, 1} \right)^{z/z_1}.
\end{equation}

$\alpha_{i, 0}$ describes the galaxy population at redshift ${z = 0}$, $\alpha_{i, 1}$ at redshift ${z = z_1 > 0}$. Eq. \eqref{eq:template-coeff-evolution} allows galaxies to smoothly transition from $\alpha_{i, 1}$ to $\alpha_{i, 0}$. Furthermore, we use different Dirichlet distributions for blue and red galaxies. Depending on whether a galaxy originated from the luminosity function for blue galaxies or from the one for red galaxies, its spectrum is sampled accordingly.

\subsubsection{Apparent magnitudes}
\label{subsubsec:gal-app-mag}

To calculate magnitudes in arbitrary filter bands, we use the AB magnitude system \cite{Oke1968, Blanton2006} which relates the magnitude $\text{mag}_x$ in the filter band $x$ to the spectral energy distribution $f(\lambda)$ via

\begin{equation}
\label{eq:mag-def}
\text{mag}_x = - 2.5 \log_{10} \left[ \frac{\int\limits_0^\infty d\lambda \, \lambda f(\lambda) R_x(\lambda)}{\int\limits_0^\infty d\lambda \, \lambda f^\text{AB}(\lambda) R_x(\lambda)} \right].
\end{equation}

$f^\text{AB}$ is the spectral energy distribution of the standard source of the AB magnitude system and $R_x$ the filter response. To calculate absolute magnitudes from eq. \eqref{eq:mag-def}, the rest-frame spectral energy distribution $f_\text{e}(\lambda)$ is integrated with respect to the rest-frame wavelength $\lambda_\text{e}$. To obtain apparent magnitudes, the observed spectral energy distribution $f_\text{o}$ has to be integrated using the observed wavelength $\lambda_\text{o}$. $f_\text{e}$ is linked to $f_\text{o}$ via (\cite{Hogg2002})

\begin{equation}
f_\text{o}(\lambda_\text{o}) = \left( \frac{10\,\text{pc}}{d_L} \right)^2 \frac{f_\text{e}(\lambda_\text{e})}{1+z}, 
\end{equation}

where $d_L$ is the luminosity distance corresponding to the redshift of the galaxy. Furthermore, we include galactic extinction, which reddens the spectrum. This can be accounted for by modifying $f_\text{o}(\lambda_\text{o})$ according to

\begin{equation}
\label{eq:extinction}
f_\text{o}(\lambda_\text{o}) \to 10^{-\frac{2}{5} A_{\lambda_\text{o}}} \, f_\text{o}(\lambda_\text{o}).
\end{equation}

The wavelength-dependent extinction coefficient $A_\lambda$ varies with the line-of-sight and is generally larger for smaller wavelengths (see appendix \ref{sec:extinction} for more details). Hence, the apparent magnitudes assigned to a galaxy are computed via

\begin{equation}
\text{mag}_x = - 2.5 \log_{10} \left[ \frac{\left( \frac{10\,\text{pc}}{d_L} \right)^2}{1+z} \frac{\int\limits_0^\infty d\lambda_\text{o} \, \lambda_\text{o} 10^{-\frac{2}{5} A_{\lambda_\text{o}}} f_\text{e}\left(\frac{\lambda_\text{o}}{1+z}\right) R_x(\lambda_\text{o})}{\int\limits_0^\infty d\lambda_\text{o} \, \lambda_\text{o} f^\text{AB}(\lambda_\text{o}) R_x(\lambda_\text{o})} \right].
\end{equation}

\subsubsection{Sizes}
\label{subsubsec:gal-size}

The distribution of intrinsic sizes of galaxies is generally well described by a log-normal distribution (\cite{Shen2003} and references therein). The model we employ characterizes the physical radius $r^\text{phys}_{50}$ of a galaxy as being distributed log-normally with mean $\mu_\text{phys}$ and standard deviation $\sigma_\text{phys}$. We use a fixed $\sigma_\text{phys}$ for all galaxies and  a variable $\mu_\text{phys}$, which depends linearly on the absolute magnitude $M$ assigned to the galaxy by the luminosity function:

\begin{equation}
\mu_\text{phys}(M) = a_\mu \, M + b_\mu,
\end{equation}

where $a_\mu$ is the slope and $b_\mu$ the intercept of the model. This particular functional form is empirically motivated (see section \ref{subsubsec:gal-size-prior}). To transform $r^\text{phys}_{50}$ into an angular size $r^\text{ang}_{50}$, which is the apparent size of an object on the sky, we use the angular diameter distance $d_A$. This redshift-dependent cosmological distance measure is defined as the ratio of an object's physical size and its angular size measured in radians \cite{Hogg1999}:

\begin{equation}
d_A = \frac{r^\text{phys}_{50}}{r^\text{ang}_{50}}.
\end{equation}

Hence, assigning an intrinsic size to a galaxy is done in two steps: First, we draw $r^\text{phys}_{50}$ from a log-normal distribution whose mean depends on the absolute magnitude of the galaxy. We then use the redshift at which the galaxy is located to compute $d_A$, which allows us to transform from physical size to angular size.

\subsection{Star population}

To generate the stars rendered on an image, we use the Besançon model \cite{Robin2003} which is based on stellar population synthesis. The code can be publicly executed using the website of the model\footnote{\url{http://model.obs-besancon.fr}}. We synthesize the stellar population corresponding to a 5$\,$deg$^2$ patch centered on the COSMOS field, which yields a catalog of stars for this region of the sky. To simulate apparent magnitudes in multiple filter bands, we set the model to use the CFHT-MegaCam photometric system and use the corresponding $g$-, $r$- $i$-, $z$-bands as $g^+$-, $r^+$- $i^+$-, $z^+$-bands. To render a single Subaru tile, we subsample from this catalog according to the area covered by the image.

\section{Test case}
\label{sec:test-case}

In this section, we describe the test case that we use to demonstrate our method. We first describe the combination of the 4-band Subaru imaging data and the complementary VVDS spectroscopic survey that is used, the target sample for which we measure $n(z)$ and the 36-band COSMOS catalog against which we verify our results. 

\subsection{Data}

\subsubsection{Subaru}
\label{subsec:subaru-data}

We use publicly available\footnote{\url{http://cosmos.astro.caltech.edu}} images taken by the Suprime-Cam instrument on the Subaru Telescope in the COSMOS field \cite{Capak2007, Taniguchi2007}. We use images taken in the four broadbands $g^+$, $r^+$, $i^+$ and $z^+$, which were PSF-homogenized, i.e. smoothed with a Gaussian kernel to achieve fixed PSF sizes. After the homogenization, the PSF size varies between $0.95''$ and $1.58''$, the pixel scale is $0.15''$. In total we use images covering an area of $1.86\,$deg$^2$. Details of the tiles we use can be found in appendix \ref{sec:used-subaru-tiles}.

To produce catalogs of detected objects from the images, we run \textsc{Source Extractor} (\textsc{SExtractor}, \cite{Bertin1996}). This software detects objects and measures various properties such as position, magnitude, size etc. We apply \textsc{SExtractor} in dual-image mode, where the detection of sources is done on one image and the measurements are performed on a potentially different, but astrometrically aligned image. This allows us to measure fluxes of the same object in multiple filter bands. We always use the $r^+$-band as the reference image for detection. The detailed configuration we use to apply \textsc{SExtractor} can be found in appendix \ref{sec:sextractor-config}.

\subsubsection{VVDS}
\label{subsec:vvds}

The VIMOS VLT Deep Survey (VVDS, \cite{LeFevre2005, Garilli2008, LeFevre2013}) is a spectroscopic redshift survey that is divided into three sub-surveys: VVDS-Wide, VVDS-Deep and VVDS-Ultra-Deep. VVDS is magnitude-limited in the $i$-band, covering the ranges $17.5 \leq \text{mag}_i \leq 22.5$ (Wide), $17.5 \leq \text{mag}_i \leq 24$ (Deep) and $23 \leq \text{mag}_i \leq 24.75$ (Ultra-Deep). We use publicly accessible\footnote{\url{http://cesam.lam.fr/vvds/}} redshifts and $i$-band magnitudes from VVDS-Wide and VVDS-Deep to provide two complete reference samples that have secure redshifts. To ensure the latter, we only use galaxies with redshift flags 2, 3 or 4, meaning that the probability of the measured redshift being correct is at least $75\%$.  This additional information effectively tightens the constraints we obtain on $n(z)$. For details of how we incorporate the VVDS data into the ABC analysis, see section \ref{subsec:distance-metrics}.

\subsection{Target sample}
\label{subsec:target-sample}
\begin{figure}[tbp]
\centering
\includegraphics[width=\textwidth]{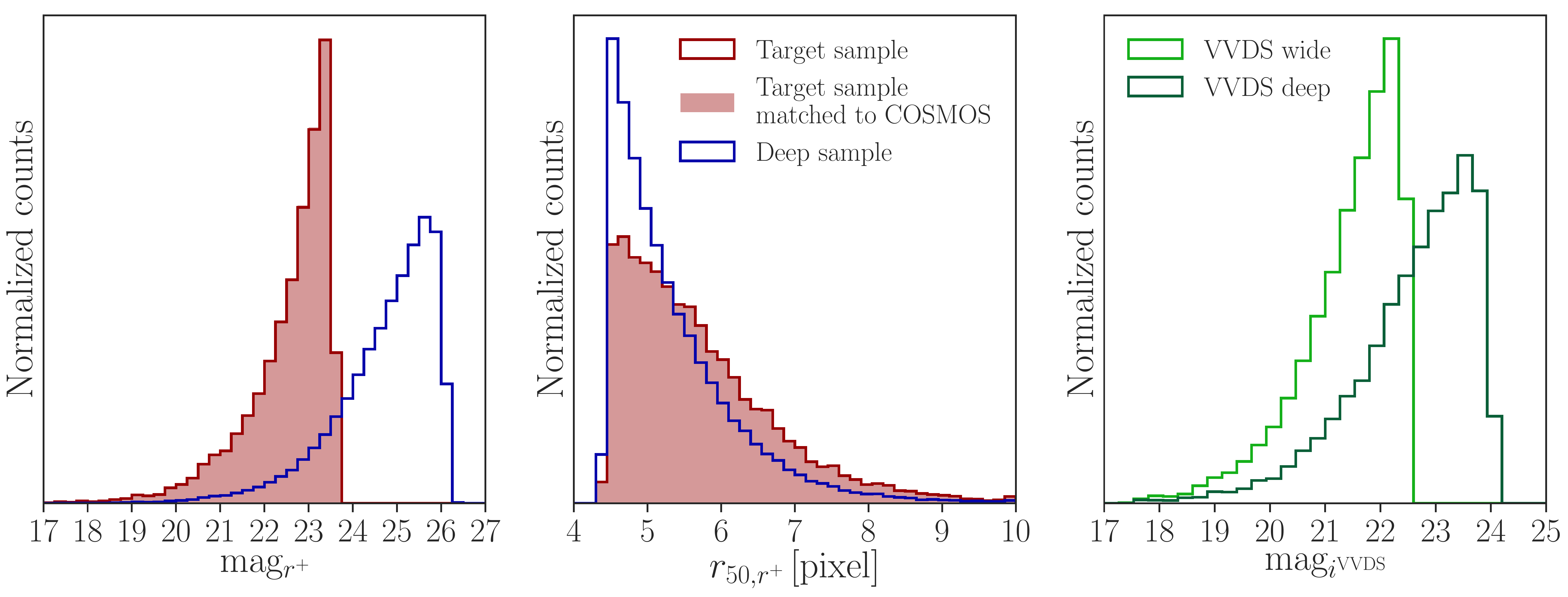}
\caption{Distributions of magnitudes (left) and sizes (middle) for the entire target sample (section \ref{subsec:target-sample}), for those galaxies in the target sample that could be matched to the COSMOS2015 catalog (section \ref{subsec:cosmos-catalog}) as well as for the deep sample (used to compute distance metrics in the ABC analysis, see section \ref{subsec:distance-metrics}). We show magnitudes and sizes determined by \textsc{SExtractor} (output columns \texttt{MAG\_AUTO} and \texttt{FLUX\_RADIUS}, both measured in the $r^+$-band). The right panel displays the $i$-band magnitudes of the VVDS Wide and Deep sample (provided by the VVDS collaboration) that we use to compute distance metrics during the ABC analysis.}
\label{fig:subaru-samples-mag-size-vvds-mag}
\end{figure}

We aim to estimate $n(z)$ for a specific sample of galaxies that we call the target sample. This target sample is designed to mimic a shear sample for current wide-field surveys such as DES. Specifically, we apply two selection cuts on the signal-to-noise ratio and the other on the ratio of an object's size to the PSF size. The signal-to-noise cut ($> 55$) was chosen such that the magnitude distribution of the target sample is similar to that of the im3shape catalog \cite{Bonnett2016}, one of the DES science verification weak lensing catalogs. The cut on the size is chosen such that the size ratio of galaxies to PSF (using half-light radius) is greater than 1.15, which is typical for weak lensing surveys to ensure that galaxies are resolved. The resulting distribution of magnitudes and sizes of the target sample is displayed in figure \ref{fig:subaru-samples-mag-size-vvds-mag}.

\subsection{Verification data}
\label{subsec:cosmos-catalog}

The Cosmic Evolution Survey (COSMOS, \cite{Scoville2007}) is designed to study the formation and evolution of galaxies up to redshifts of $z \sim 6$. It covers an area of $2\,$deg$^2$ and uses both space- and ground-based telescopes. We use the publicly available\footnote{\url{http://cosmos.astro.caltech.edu}} COSMOS2015 catalog described in \cite{Laigle2016} to test our results on $n(z)$. This catalog contains high-quality photometric redshifts based on $36$ filter bands to a $3 \, \sigma$-depth of more than $26$ magnitudes in the $r^+$-band. By matching objects in our target sample to objects listed in the COSMOS2015 catalog, we are able to obtain an estimate of the $n(z)$ of the target sample we are aiming at constraining. We cross-identify objects based on position and $r^+$-band magnitude using the matching criteria of maximum distance of $1.5''$ and a maximum magnitude difference of $1$.

Not all galaxies in the target sample can be matched with objects in the COSMOS2015 catalog according the matching criteria explained in above. Thus, our verification sample is not identical to our target sample. However, the matching does not induce any significant bias and the two samples are statistically virtually identical. This can also be seen in figure \ref{fig:subaru-samples-mag-size-vvds-mag}, where the entire target sample is compared to the subset of the target sample that can be cross-identified with objects in the COSMOS2015 catalog.

\section{Implementation}
\label{sec:adjusting-ufig}

We now describe how we apply our method described above to the test case. We first describe how we adjust instrumental and observational parameters. We then show how we apply the ABC analysis to adjust the galaxy population parameters and thus constrain $n(z)$. We also describe the distance metrics we use to compare the simulations to the real data. Note that for the test case, we simulate all Subaru tiles that we use for each set of model parameters.

\subsection{Instrumental and observational parameters}
\label{subsec:instrument-ob-par}

The images generated by \textsc{UFig} depend on a number of instrumental and observational parameters, such as the gain of the detectors, the size of the PSF and the background level. These parameters need to be adjusted in order to obtain images that match real Subaru data. Concerning the instrumental parameters, we either read them off from the image headers or we obtained them through private communication\footnote{Dr. Peter L. Capak, Spitzer Science Center, California Institute of Technology}. The background level and the size of the PSF before smoothing was applied are obtained directly from the data, separately for each tile.

We estimate the background level by first applying the segmentation map produced by \textsc{SExtractor}. Subsequently, we sigma-clip the remaining pixels to exclude possible object pixels leaking into the sample classified as background by \textsc{SExtractor}. The surviving pixels are then used to estimate the mean and the standard deviation of the background.

The size of the PSF before smoothing was applied is estimated by identifying stars detected by \textsc{SExtractor} on the unconvolved images and using their measured size as an estimator for the PSF. We identify stars between $\sim 21$ and $\sim 24$ magnitudes as the smallest objects within this magnitude interval. We only use stars within this range to avoid detector non-linearities and saturation effects (at the bright end) as well as low signal-to-noise measurements (at the faint end), both of which can result in biased size estimates.

To properly account for the PSF homogenization that was performed on the Subaru data, we first simulate unconvolved images and then smooth them with a spherically symmetric Gaussian kernel. The size of the kernel is computed using the size of the PSF before homogenization and the target size, which is a fixed number per filter band.

\subsection{Variable parameters and priors}
\label{subsec:var-par-prior}

Using ABC, we explore the space of those \textsc{UFig} input parameters $n(z)$ is most sensitive to. In the following, we list these parameters and specify the priors we use. The combined prior is a product of the individual priors described below. The corresponding parameter space has $31$ dimensions.

\subsubsection{Galaxy luminosity function} 
\label{subsubsec:gal-lum-fct-prior}

\begin{figure}[tbp]
\centering
\includegraphics[width=\textwidth]{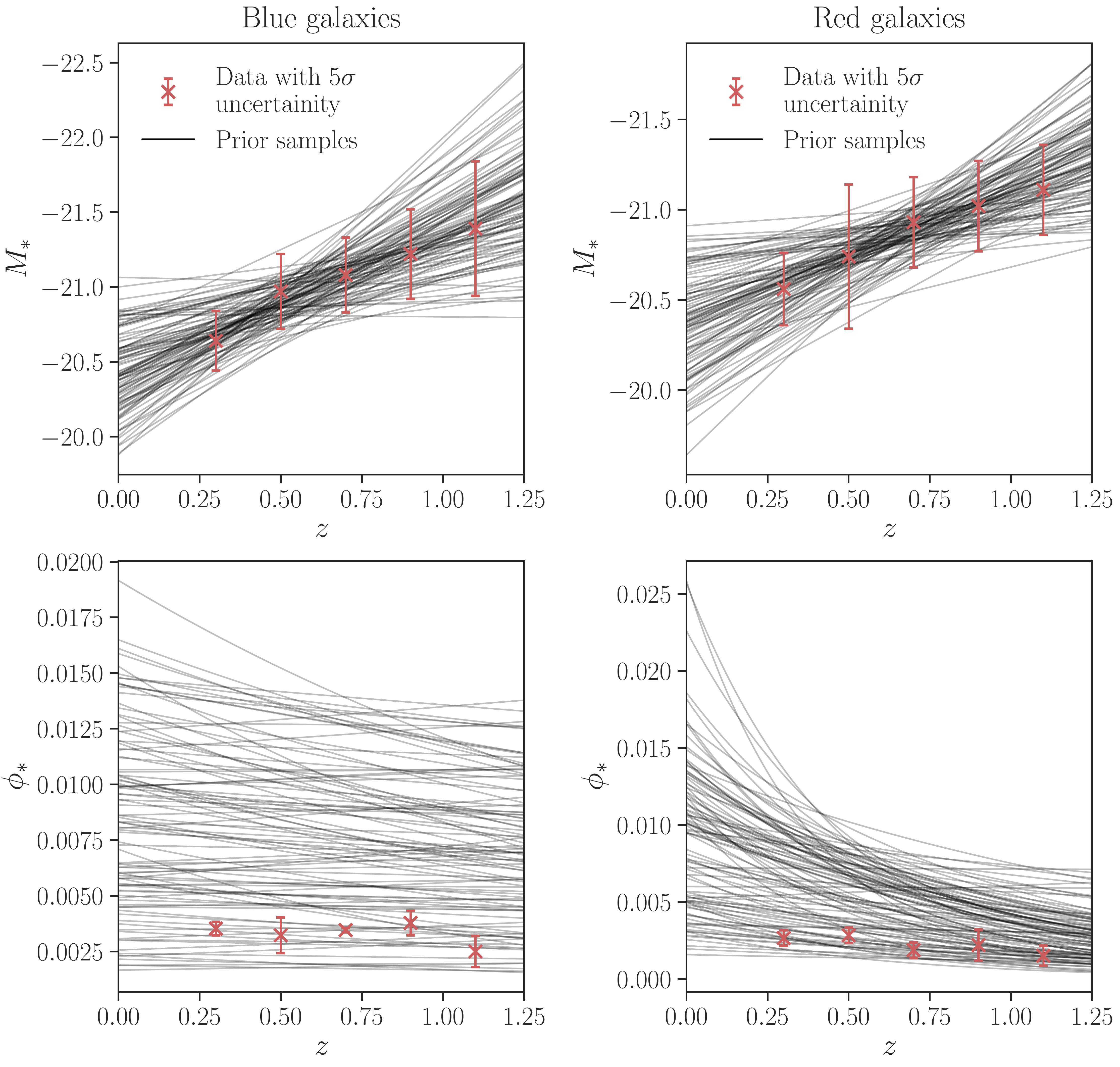}
\caption{Evolution of $M_*$ and $\phi_*$ with redshift for blue galaxies (left side) and red galaxies (right side). The datapoints are taken from \cite{Beare2015}, while we increased the uncertainties by a factor of $5$ for robustness to derive the prior. The evolution according to samples from the prior that will be used in the ABC analysis is shown as the thin grey lines. As described in the text and as can be seen in the lower two panels, we artificially increased the prior range on $b_\phi$.}
\label{fig:lum-fct-5sigma-data-prior}
\end{figure}

We vary the parameters controlling the redshift evolution of the galaxy luminosity function (see eqs. (\ref{eq:lum-fct-redshift-evolution-m*}, \ref{eq:lum-fct-redshift-evolution-phi*})). To obtain a prior on the corresponding eight-dimensional space (four parameters $a_M$, $b_M$, $a_\phi$, $b_\phi$ for blue and red galaxies, respectively), we use measurements of $\phi$ at several redshifts in the $B$-band presented in \cite{Beare2015}, as shown in figure \ref{fig:lum-fct-5sigma-data-prior}. Blue and red galaxies were treated separately in this work. We set the parameter $\alpha$ to $-1.3$ for blue and to $-0.5$ for red galaxies, which are the values determined by \cite{Beare2015}.

Assuming that the uncertainties stated by \cite{Beare2015} are Gaussian, we are able to write down a likelihood for a a given set of parameters $a_M$, $b_M$, $a_\phi$, $b_\phi$ (with different values for the blue and the red luminosity function). We sample the corresponding space via Markov chain Monte Carlo (MCMC) sampling using the \texttt{CosmoHammer} package \cite{Akeret2013}. The resulting samples are employed as input for the ABC analysis. To obtain a conservative prior, we multiply the uncertainties stated in \cite{Beare2015} by a factor of 5. Furthermore, we enlarge the prior volume for $b_\phi$ after sampling with \texttt{CosmoHammer} by multiplying with a random number between $0.5$ and $4$ (separately for blue and red galaxies). This is necessary in order to be able to fit the data and obtain simulations that are statistically close to the Subaru data (see section \ref{subsec:post-sim-sub-data}). The resulting prior is also shown in figure \ref{fig:lum-fct-5sigma-data-prior}.

\subsubsection{Intrinsic galaxy sizes} 
\label{subsubsec:gal-size-prior}

The sizes of galaxies are set by three parameters ($a_\mu$, $b_\mu$, $\sigma_\text{phys}$) in \textsc{UFig} (see section \ref{subsubsec:gal-size}). To obtain a prior on these parameters, we use data from the third GRavitational lEnsing Accuracy Testing challenge (GREAT3, \cite{Mandelbaum2014}). This international project was aimed at testing the accuracy of shear measurement algorithms based on image simulations. To enable the realistic modeling of images of single galaxies, the GREAT3 team performed fits of Sérsic galaxy profiles to images taken by the \textit{Hubble Space Telescope} as part of COSMOS. This yielded publicly accessible\footnote{\url{http://great3.jb.man.ac.uk/leaderboard/data}} distributions of parameters affecting the Sérsic profile, such as the half-light radius $r_{50}$.

\begin{figure}[tbp]
\centering
\includegraphics[width=0.6\textwidth]{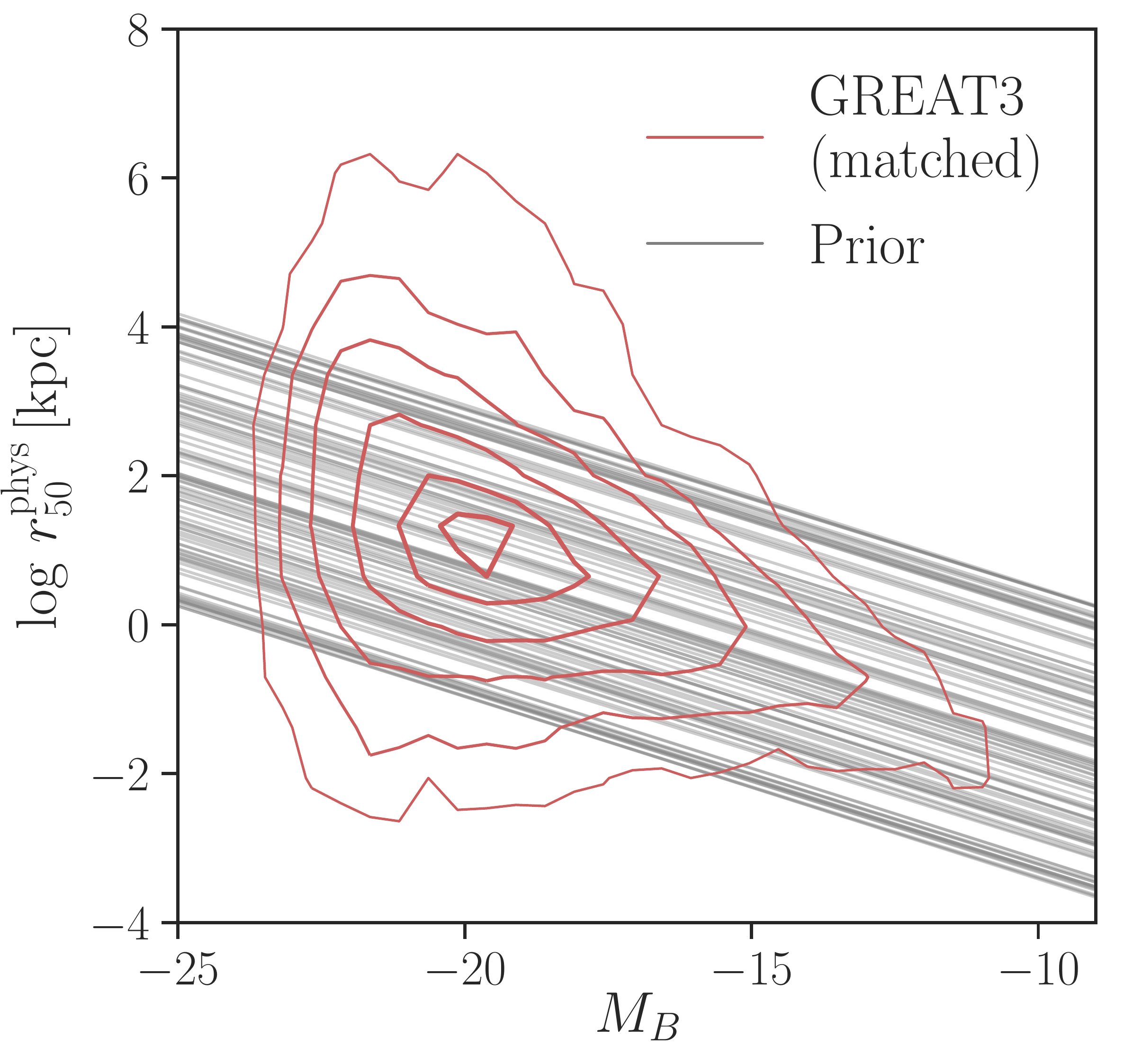}
\caption{Relation between the absolute magnitude and the logarithm of the physical size of galaxies. The contours show the data from the fits performed by the GREAT3 collaboration, whereby we matched the galaxies they fitted to galaxies in the COSMOS2015 catalog in order to obtain absolute magnitudes and redshifts. The contours start from $0.5 \, \sigma$ and increase in steps of $0.5 \, \sigma$. They grey lines represent samples from the prior that we will use in the ABC analysis. They correspond to different values of the slope $a_\mu$ and the intercept $b_\mu$ used in our model for the physical size of galaxies (see section \ref{subsubsec:gal-size}).}
\label{fig:abs-mag-log-size-prior}
\end{figure}

We use this information in conjunction with redshifts and absolute magnitudes from COSMOS to obtain a prior on the parameters controlling our model of the physical sizes of galaxies. For this purpose, we match fitted galaxies to the COSMOS2015 catalog. The matching is based on position and redshift, since those are the parameters contained in both catalogs. The criteria for a match are a maximum distance of $1.5''$ and a maximum redshift difference of $0.05$. We then convert the sizes measured by the GREAT3 collaboration to physical sizes (see section \ref{subsubsec:gal-size}). The resulting relation between absolute magnitude and physical size is visualized in figure \ref{fig:abs-mag-log-size-prior}.

To obtain a prior for ABC, we sample the slope, the intercept and the standard deviation using \texttt{CosmoHammer}. The likelihood function is in this case directly given by our size model evaluated at the points where GREAT3 data is available. However, the resulting prior is too restrictive to allow for a good fit to the Subaru data. Out of the three parameters affecting our size model, the ABC analysis is most sensitive to the value of the intercept $b_\mu$. Relaxing the prior on this parameter to be uniform between $-1$ and $3$ allows us to find parameter configurations that result in simulated images which are statistically close to the Subaru data. Figure \ref{fig:abs-mag-log-size-prior} visualizes the prior on the parameters affecting the model for the physical size of galaxies.

\subsubsection{Galaxy spectra}
\label{subsubsec:gal-spectrum-prior}

We also vary the Dirichlet distributions from which galaxy rest frame spectra are sampled (see section \ref{subsubsec:gal-spectrum-model}). The corresponding space has 10 dimensions for blue and red galaxies, respectively, since we vary both $\alpha_{i, 0}$ (at redshift ${z = 0}$) and $\alpha_{i, 1}$ (at redshift ${z = z_1}$), with ${i = 1,2,...,5}$. We choose ${z_1 = 1}$, since this value is approximately in the center of the range of redshifts we are sensitive to in this paper. To vary the Dirichlet parameters during the ABC analysis, we require a prior that allows us to efficiently explore the space of possible combinations of $\alpha_i$. A way to construct such a prior is to use a flat Dirichlet distribution of order 5. This distribution is characterized by equal weights on all dimensions, i.e. by ${\alpha_i = 1}$. This means that the prior does not favor any particular template spectrum but instead gives equal weight to all of them. 

Furthermore, we assume a uniform prior ranging from $5$ to $15$ on the sum of the Dirichlet parameters. This effectively changes the standard deviation of the corresponding Dirichlet distribution, allowing the ABC analysis to explore a greater variety of models. The range of the prior on the sum of the Dirichlet parameters was chosen according to fits of Dirichlet distributions to real data that we performed. Specifically, we fitted Dirichlet distributions of order five to samples of template coefficients derived from SDSS data (see appendix \ref{sec:nyu-vagc} for details). In total, we use 4 Dirichlet priors of order 5 in conjunction with 4 corresponding uniform priors.

\subsection{Distance metrics}
\label{subsec:distance-metrics}

We now describe the distance metrics we use in the ABC analysis to constrain $n(z)$. All of them operate on the \textsc{SExtractor} catalogs, where we have run \textsc{SExtractor} on the simulated images using exactly the same configuration as was used to reduce the data. We chose distance metrics that test the basic properties of the samples such as number counts, the distribution of magnitudes and sizes and the distribution of galaxy colors. We also include distance metrics that use spectroscopic VVDS data. In total, we use five distance metrics. 

To compute distance metrics that compare the simulations to the Subaru data, we use a sample of galaxies that is deeper than the target sample. We call this the deep sample. It includes all galaxies with a signal-to-noise ratio greater than 15 and a ratio of galaxy to PSF size greater than 1.15. The rationale behind this approach is that a deeper sample contains more objects, resulting in tighter constraints on the model parameters. Figure \ref{fig:subaru-samples-mag-size-vvds-mag} displays the distributions of magnitudes and sizes for the deep sample. Furthermore, we use the VVDS Wide and Deep samples described in section \ref{subsec:vvds} to compute distance metrics. The magnitude distributions of these two samples provided by the VVDS collaboration are displayed in the right panel of figure \ref{fig:subaru-samples-mag-size-vvds-mag}.

\paragraph*{Histogram distance}

To assess the difference $d_\text{hist}$ between two equally binned histograms $h_1$ and $h_2$, we sum up the absolute values of the differences between the counts in the bins:

\begin{equation}
d_\text{hist} = \sum_i \left| h_{1, i} - h_{2, i} \right|,
\end{equation}

where $h_{1, i}$ and $h_{2, i}$ are the counts in the $i$-th bin of the first and the second histogram, respectively. By construction, this distance metric is sensitive to both the overall number counts as well as the shapes of the two histograms. We apply it to histograms of magnitudes as well as sizes measured in the $r^+$-band of all objects in the deep galaxy sample. To evaluate $d_\text{hist}$ for one set of \textsc{UFig} input parameters on multiple Subaru tiles, we use averaged histograms, i.e., we stack the individual histograms and average the bin entries.

\paragraph*{MMD distance} 

The Maximum Mean Discrepancy (MMD, \cite{Gretton2012}) distance $d_\text{MMD}$ is used to measure the difference between the probability distributions underlying two given samples $x$ and $y$. It is calculated via

\begin{equation}
d_\text{MMD} = \frac{1}{N(N-1)} \sum_{\substack{i, j \\ i \neq j}} k(x_i, x_j) + k(y_i, y_j) - k(x_i, y_j) - k(y_i, x_j),
\end{equation}

where $x_i$ and $y_i$ are the $i$-th element of the two samples, respectively, and $N$ denotes the size of $x$ and $y$. The kernel function $k$ is given by a Gaussian kernel of a pre-defined width $\sigma$:

\begin{equation}
k(x_i, y_j) = \exp \left( -\frac{\left\Vert x_i - y_j \right\Vert^2}{2 \sigma} \right),
\end{equation}

where $||\cdot||$ denotes the Euclidian norm. $\sigma$ is a free parameter that we choose according to the median Euclidian distance between the elements of $x$ and $y$ in the case where $x$ and $y$ are drawn from the same probability distribution, as suggested in \cite{Gretton2012}. To apply the MMD distance, we always first transform $x$ and $y$ along each dimension to have approximately zero mean and a standard deviation of $1$. We calculate the following MMD distances:

\begin{enumerate}

\item We compute an eight-dimensional MMD distance between the measured magnitudes and sizes in the four filter bands using all galaxies in the deep sample. Here, we use $\sigma = 2.081$. This distance metric is sensitive to magnitudes, sizes and colors.

\item We also use the MMD distance in combination with the spectroscopic data from VVDS (see section \ref{subsec:vvds}). That is, we compute two-dimensional MMD distances between $i$-band magnitudes and redshifts from VVDS and (measured) $i^+$-band magnitudes and redshifts from \textsc{UFig}. Specifically, we apply the VVDS-Wide and VVDS-Deep magnitude cuts to all galaxies detected on the simulated images and use the resulting samples to compare to the VVDS data. We use $\sigma = 1.065$ (Wide) and $\sigma = 0.941$ (Deep).

\end{enumerate}

To compute $d_\text{MMD}$ for one \textsc{UFig} configuration evaluated on multiple Subaru tiles, we first compute the distances tile by tile and then average.

\section{Results}
\label{sec:results}

In this section, we present our results from applying our method to the test case. We drew a total of $140\,000$ samples from the prior described in section \ref{subsec:var-par-prior} (whereby a "sample" refers to one model parameter configuration), evaluated the model for these points and calculated the distances listed in section \ref{subsec:distance-metrics}. From these samples, we obtain a posterior by thresholding on all distance metrics combined. Specifically, to obtain the $N$ best samples, we find the fraction $q \in [0, 1]$ such that there are $N$ samples whose associated distance metrics are all smaller than the $q$-th percentiles of the distance metrics of all prior samples. In what follows, we always use the posterior consisting of $N = 150$ samples.

\subsection{Comparison of adjusted simulations and data}
\label{subsec:post-sim-sub-data}

\begin{figure}[tbp]
\centering
\includegraphics[width=\textwidth]{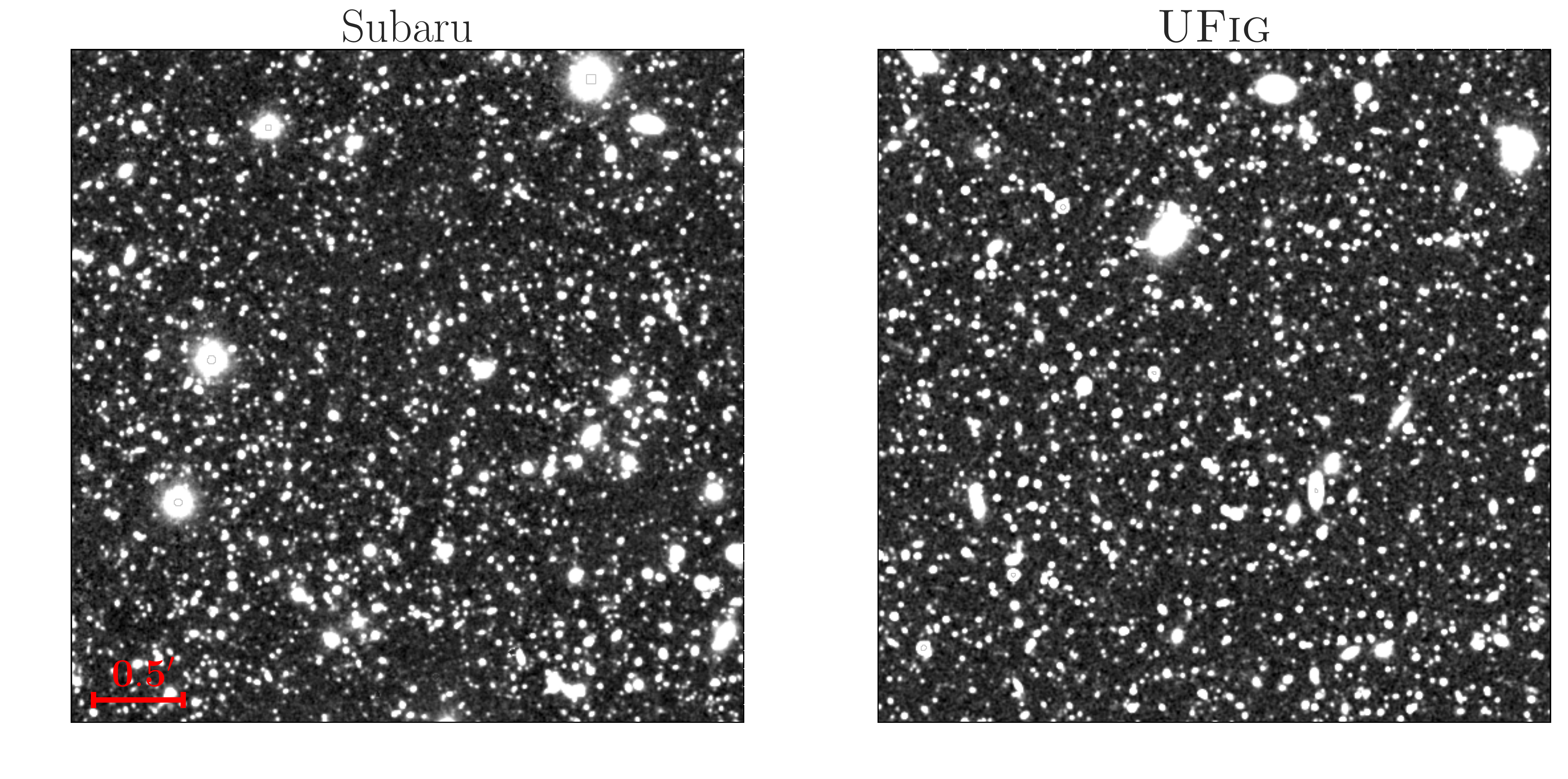}
\caption{Visual comparison of Subaru data (left) and a simulated image (right) corresponding to one of the posterior samples. We show cutouts from tile no. 78 in the $r^+$-band, covering an area of $\sim 14'^2$. The red reference bar denotes an angular distance of $0.5'$. The color scales are the same on the left and on the right.}
\label{fig:subaru-ufig-img}
\end{figure}

We first compare the image simulations associated with the posterior samples to Subaru data. Figure \ref{fig:subaru-ufig-img} shows a cutout from a Subaru tile and a cutout from an image simulated using one of the posterior parameter configurations. As can be seen on the figure, they appear similar visually.

\begin{figure}[tbp]
\centering
\includegraphics[width=\textwidth]{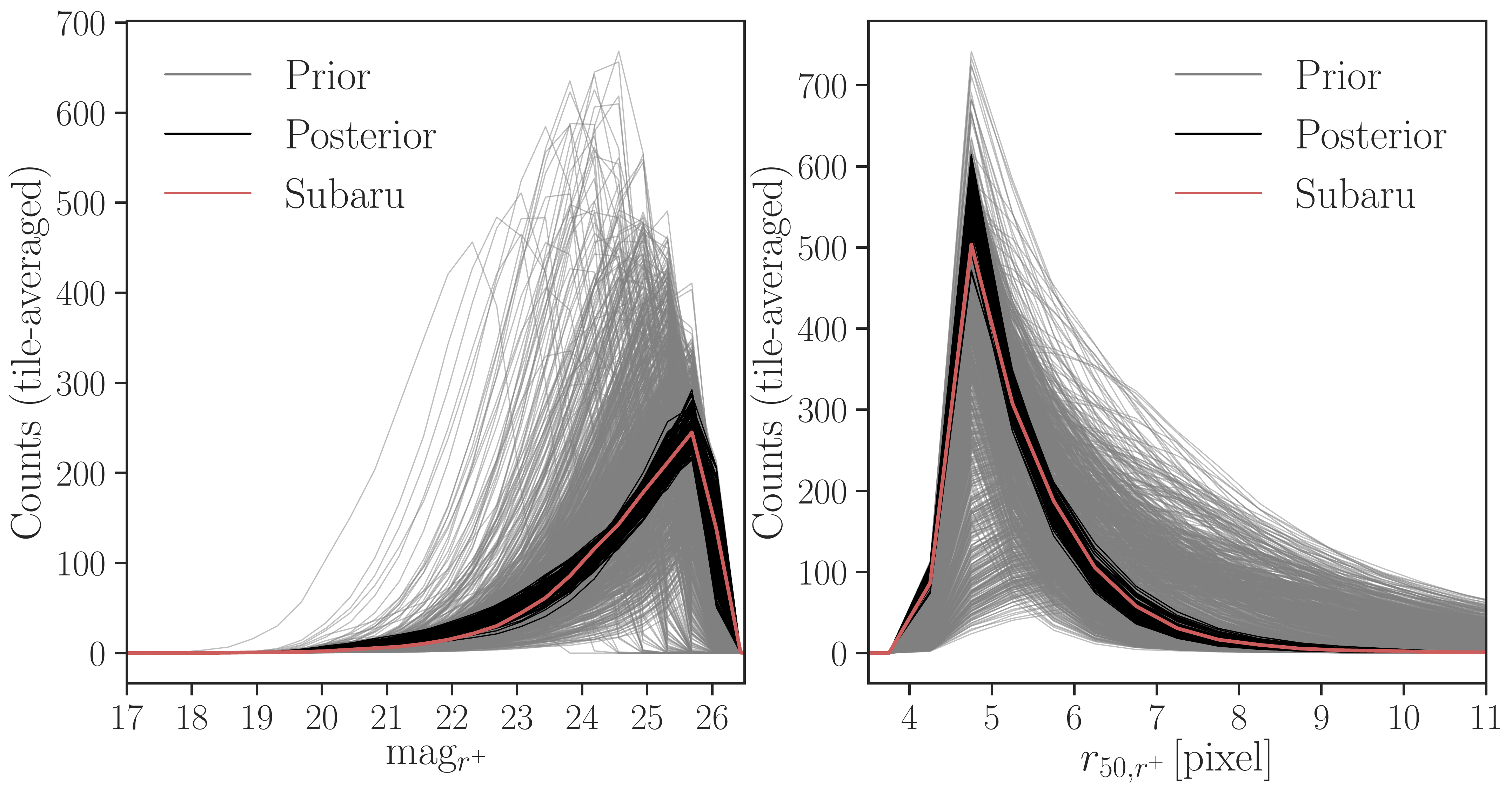}
\caption{Comparison of Subaru data and \textsc{UFig} simulations after applying ABC. The left panel shows histograms of $r^+$-magnitudes of detected objects for the data as well as for the posterior samples, averaged over all tiles that we used. Additionally, we show the same histograms for randomly chosen samples from the prior to demonstrate how ABC is able to constrain regions in parameter space using the data. The right panel shows the same in terms of measured sizes, again in the $r^+$-band. In both cases, the deep galaxy sample was used to create the histograms.}
\label{fig:mag-size-hist-post-deep}
\end{figure}

For a quantitive comparison, we consider basic statistical properties of the data and the simulations such as the number of detected galaxies, their magnitudes and sizes as well as their colors. Figure \ref{fig:mag-size-hist-post-deep} shows histograms of the measured magnitudes and sizes for all posterior samples as well as for the Subaru data, in both cases of the deep galaxy sample. The histograms correspond to all used tiles combined, i.e., we average the histograms of the individual tiles. Additionally, we also show the histograms corresponding to a number of random samples from the prior. This illustrates how ABC is able to constrain regions in parameter space which yield good fits to the data.

\begin{figure}[tbp]
\centering
\includegraphics[width=\textwidth]{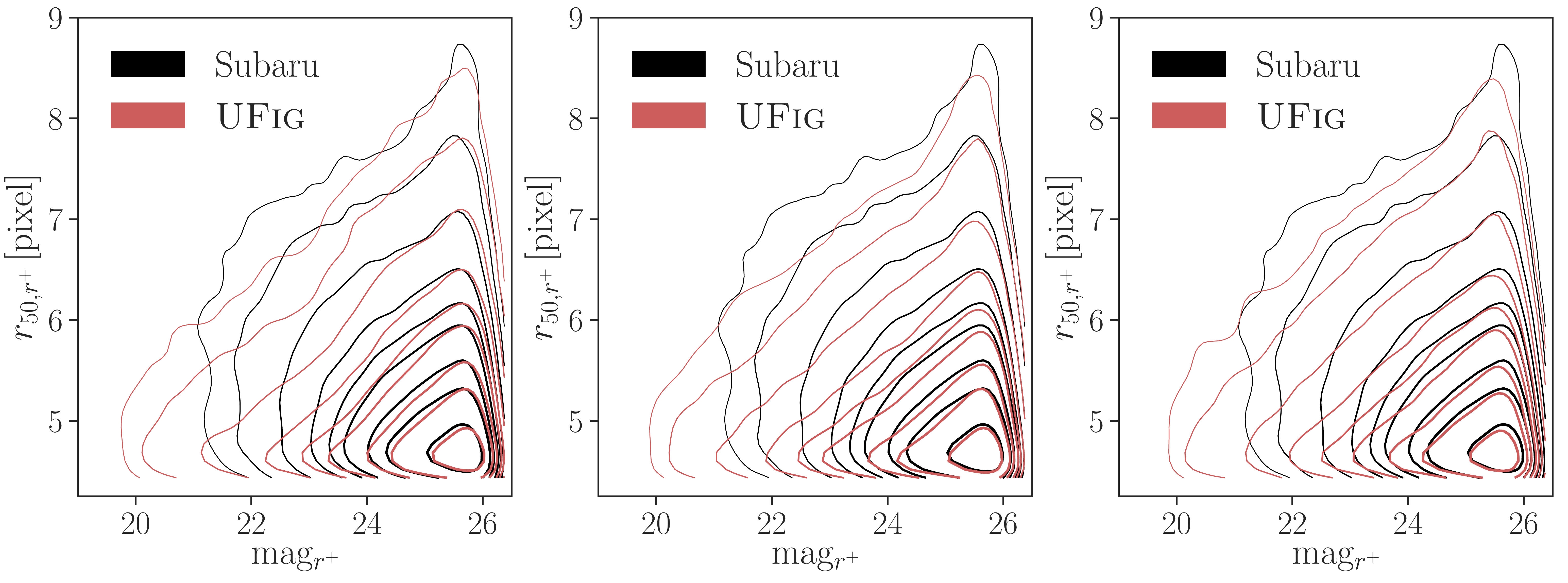}
\caption{Magnitude-size distribution of galaxies in the Subaru data and the \textsc{UFig} simulations after applying ABC for 3 posterior samples selected at random. We show magnitudes and sizes of the deep galaxy sample measured in the $r^+$-band.}
\label{fig:mag-size-plane-post-samples-deep}
\end{figure}

As figure \ref{fig:mag-size-hist-post-deep} demonstrates, our simulations match the data in terms of number counts and also in terms of magnitudes and sizes. We furthermore compare the joint distribution of magnitudes and sizes in the data and in the simulations, again using the deep galaxy sample. Figure \ref{fig:mag-size-plane-post-samples-deep} shows the magnitude-size-plane in the $r^+$-band for 3 randomly selected posterior samples as well as for the data.

\begin{figure}[tbp]
\centering
\includegraphics[width=\textwidth]{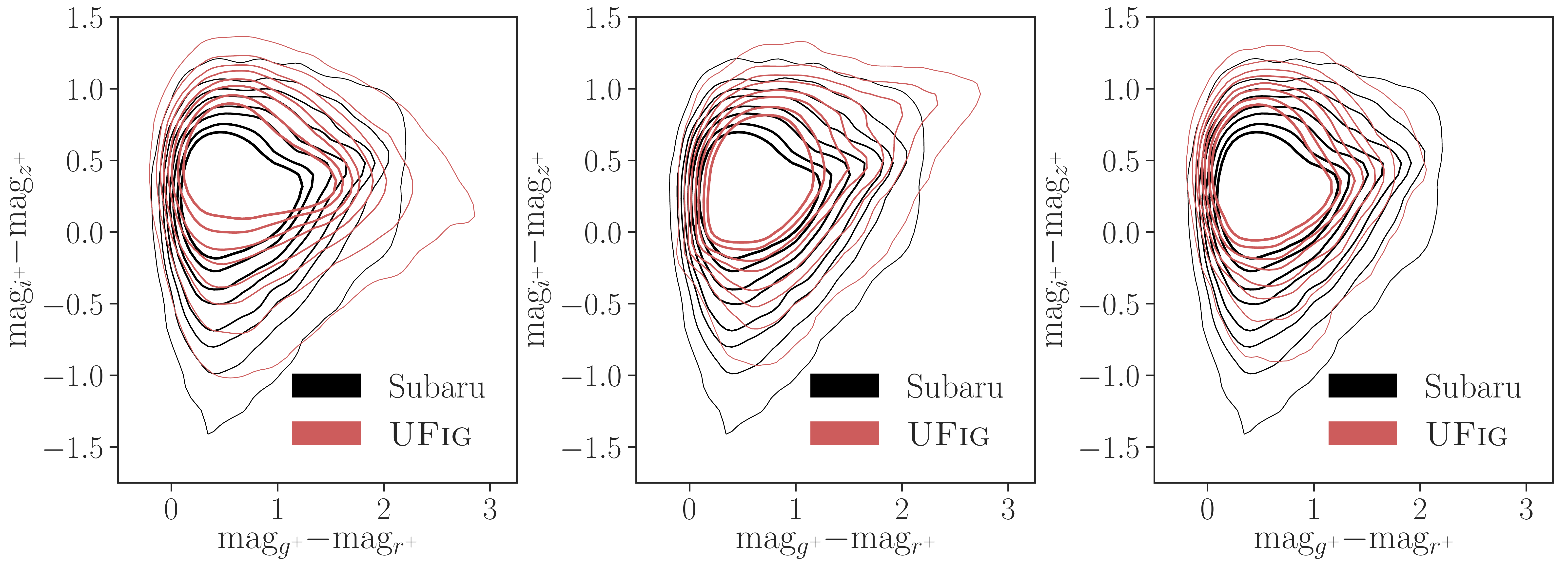}
\caption{Comparison of galaxy colors measured on Subaru data to galaxy colors measured on \textsc{UFig} images for 3 randomly selected posterior samples. We show the distribution of the color $g^+ - r^+$ vs. the color $i^+ - z^+$ of all galaxies in the deep sample.}
\label{fig:g-r-i-z-post-samples-deep}
\end{figure}

We again see that \textsc{UFig} in conjunction with ABC is able to generate images which are statistically similar to real data. So far, all comparisons were based on the $r^+$-band only. To check inter-band correlations, we compare the colors of galaxies in the simulations to the colors of galaxies measured on the data. Figure \ref{fig:g-r-i-z-post-samples-deep} shows the two-dimensional distribution of the colors $g^+ - r^+$ and $i^+ - z^+$ using the deep galaxy sample.

We conclude that ABC is able to find regions in parameter space where \textsc{UFig} produces images which are statistically close to real Subaru data. This conclusion is based on the comparison of basic properties of the simulations and the data, namely galaxy number counts, magnitudes, sizes and colors.

\subsection{\boldmath Posterior for $n(z)$}
\label{subsec:posterior-nz}

\begin{figure}[tbp]
\centering
\includegraphics[width=0.6\textwidth]{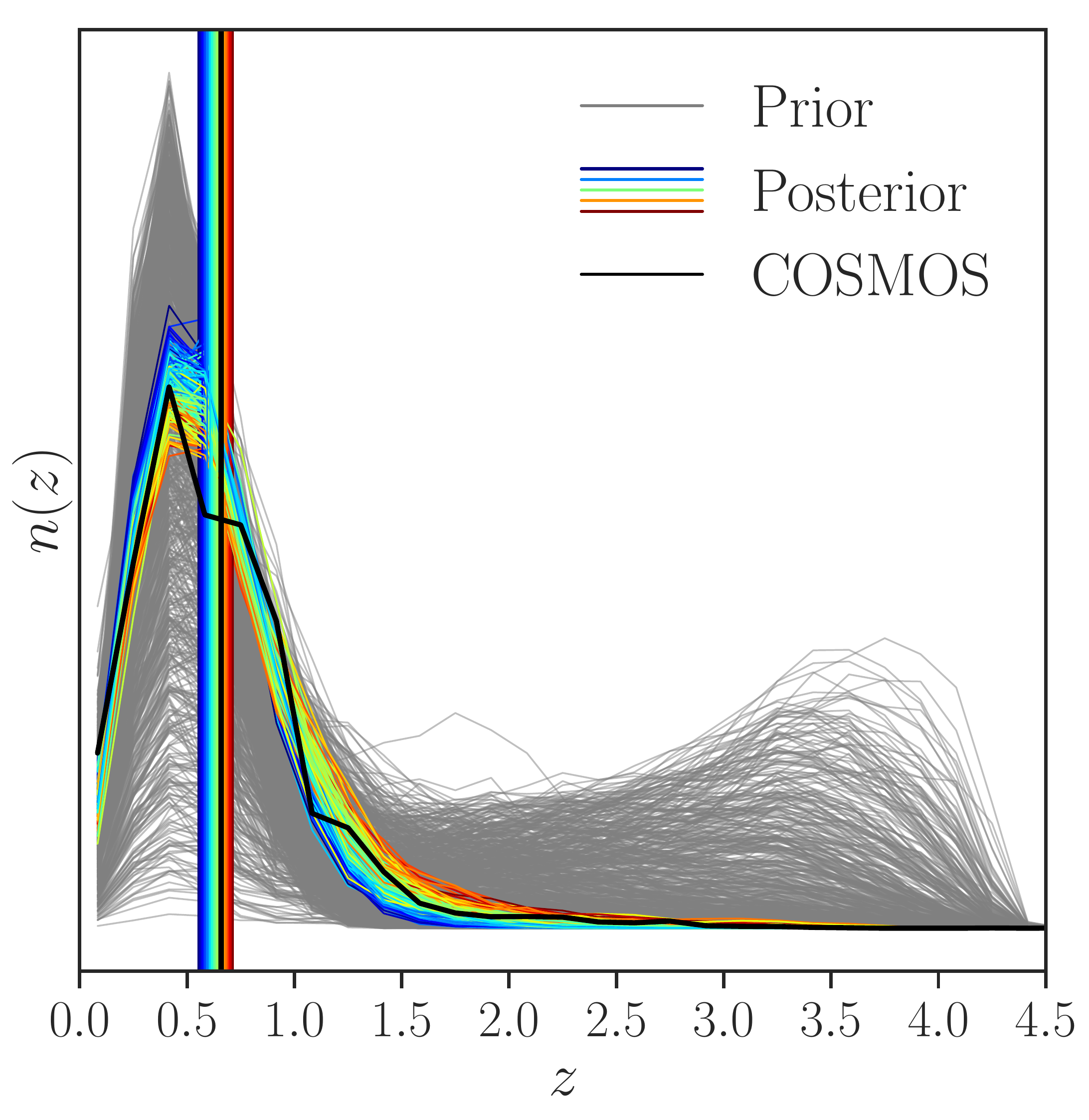}
\caption{ABC posterior for $n(z)$. The curves are color-coded according to the associated mean redshifts, which are visualized by the vertical lines. We also plot $n(z)$ for randomly selected samples from the prior to visualize how ABC is able to constrain regions in parameter space. Furthermore, we compare our posterior to the estimate from matching galaxies detected on the Subaru images to the COSMOS2015 catalog (section \ref{subsec:cosmos-catalog}). The black vertical line marks the estimate of the mean redshift of the target sample obtained from the COSMOS photometric redshifts.}
\label{fig:nz-post}
\end{figure}

After establishing that our simulated images match the Subaru data, we now present the posterior for $n(z)$ of the target sample in figure \ref{fig:nz-post}. The displayed curves are constructed using the input redshifts that \textsc{UFig} assigns to galaxies as described in section \ref{subsubsec:gal-counts-z-abs-mag} according to the posterior parameter configurations found by ABC after the selection cuts for the target sample were applied. These redshifts solely depend on the parameters of the luminosity functions and are sampled according to eq. \eqref{eq:z-mabs-dist}, jointly with absolute magnitudes. We compare our results to an estimate of $n(z)$ from the COSMOS2015 catalog. The figure shows that the former is well contained within our posterior distribution. This confirms that our method is able to provide a good measure of the $n(z)$ of cosmic shear like galaxy samples. Furthermore, we demonstrate how ABC is able to provide a proper handle on the uncertainty on $n(z)$, which can be propagated through any further analysis. We note that we do not only obtain an estimate of the uncertainty on the mean redshift, but rather a family of likely redshift distributions that quantifies the full uncertainty on $n(z)$. 

The $n(z)$ curves we infer are generally smoother than the redshift distribution estimated using the COSMOS2015 catalog. The additional features present in the latter result from non-linear small scale structure along the line-of-sight. Since our model does not include galaxy clustering, we infer $n(z)$ distributions which correspond to large scale features. Furthermore, we note that despite the small size of the COSMOS field, cosmic variance is not an issue for us when demonstrating our method, since the target and the validation sample share the same footprint.

The resulting estimate for the mean redshift of the target sample is ${0.626 \pm 0.033}$, where the central value is the average of the posterior mean redshifts and the uncertainty corresponds to $1 \, \sigma$. This can be compared to the mean redshift of $0.659$, which was estimated from the 36-band photometry in the COSMOS2015 catalog. These two results are in agreement at the $1 \, \sigma$ level. We note that our uncertainty on the mean redshift is of the same order of magnitude as the uncertainty on the mean redshift of a typical DES lensing sample. \cite{Bonnett2016} report an uncertainty of 0.01 for the mean redshift of the DES science verification weak lensing catalog, whereby the redshift estimation was performed using both template fitting and machine learning methods.

\section{Conclusions}
\label{sec:conclusions}

We presented a novel approach for determining the redshift distribution of cosmological galaxy samples. Our method relies on wide-field \textsc{UFig} image simulations to forward-model the data and on ABC to infer $n(z)$ along with its uncertainties within the \textit{MCCL} framework.

To enable the usage of \textsc{UFig} to constrain $n(z)$, we have extended its functionality to render images in arbitrary filter bands. We assign redshifts and absolute magnitudes to galaxies via luminosity functions that evolve with redshift. The evolution is parametrized based on data measured at several redshifts. Furthermore, we have implemented parametric models to assign spectra and sizes to galaxies, which allows us to vary those models during an ABC run.

We have applied our method to Subaru Suprime-Cam images of the COSMOS field, taking into account the instrumental and observational conditions. Moreover, we used spectroscopic data from VVDS as a complementary calibration dataset. By comparing the statistical properties of the simulations and the data, we have verified that our model is able to produce images that are statistically similar to the Subaru images. Applying the cuts for a cosmic shear like target sample, we derived a family of $n(z)$ distributions for the resulting acceptable models. We compared our results against photometric redshift measurements from COSMOS and found good agreement. This offers good prospects for applying our approach to large imaging survey such as DES.

\acknowledgments

This work was supported in part by grant number 200021\_169130 from the Swiss National Science Foundation. The authors thank Peter Capak for helpful discussions about the Subaru data in the COSMOS field and for providing some of the instrumental parameters for this dataset. We thank Julian Riebartsch and Lorenza Della Bruna for useful discussions.

\bibliographystyle{JHEP}
\bibliography{nz_method_subaru}

\appendix

\section{Used Subaru tiles}
\label{sec:used-subaru-tiles}

\begin{figure}[tbp]
\centering
\includegraphics[width=0.8\textwidth]{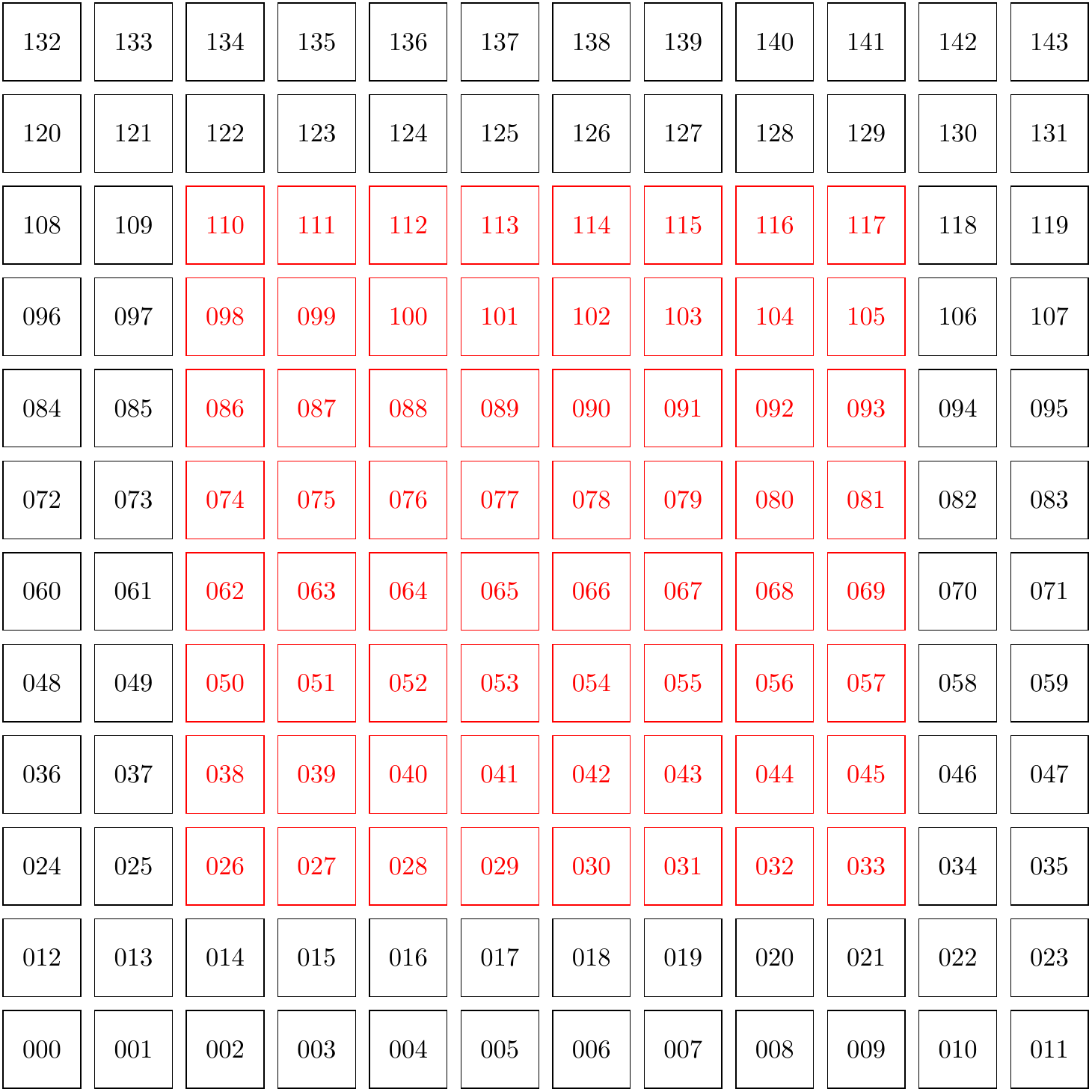}
\caption{Illustration of the Subaru tiles that we use to constrain $n(z)$. All used tiles are marked in red. We chose to use only the central tiles in order to avoid any kind of border effects.}
\label{fig:used-subaru-tiles}
\end{figure}

Figure \ref{fig:used-subaru-tiles} shows the Subaru tiles used in this paper. They are located in the COSMOS field and the data is described in \cite{Capak2007, Taniguchi2007}. The tiles we use cover an area of $1.86\,$deg$^2$. We chose to only use the central tiles to avoid border effects.

\section{\textsc{SExtractor} configuration}
\label{sec:sextractor-config}

Table \ref{tab:sextractor-config} lists the \textsc{SExtractor} configuration that was used to analyze the Subaru data and the simulated images.

\begin{table}[htb]
\centering
\begin{tabular}{| c | c |}
\hline
\textsc{SExtractor} parameter name & Value \\
\hline

\texttt{CATALOG\_TYPE} & FITS\_LDAC \\

\texttt{DETECT\_TYPE} & CCD \\    
    
\texttt{DETECT\_MINAREA} & 5 \\
       
\texttt{DETECT\_THRESH} & 1.7 \\       
   
\texttt{ANALYSIS\_THRESH} & 1.7 \\   
      
\texttt{FILTER} & Y \\      

\texttt{FILTER\_NAME} & gauss\_3.0\_5x5.conv \\

\texttt{DEBLEND\_NTHRESH} & 32 \\            

\texttt{DEBLEND\_MINCONT} & 0.005 \\      

\texttt{CLEAN} & Y \\ 
        
\texttt{CLEAN\_PARAM} & 1.0 \\      

\texttt{MASK\_TYPE} & CORRECT \\

\texttt{PHOT\_APERTURES} & 5 \\  
     
\texttt{PHOT\_AUTOPARAMS }  & 2.5, 3.5 \\

\texttt{PHOT\_FLUXFRAC}  & 0.5 \\
   
\texttt{SATUR\_LEVEL}  & tile- \& band-dependent \\

\texttt{MAG\_ZEROPOINT} & $31.4$ \\

\texttt{GAIN} & band-dependent \\

\texttt{PIXEL\_SCALE}  & $0.15$ \\

\texttt{SEEING\_FWHM}  & tile- \& band-dependent \\

\texttt{STARNNW\_NAME}  & default.nnw \\

\texttt{BACK\_SIZE} & 64 \\
            
\texttt{BACK\_FILTERSIZE} & 3 \\            

\texttt{BACKPHOTO\_TYPE} & LOCAL \\    
      
\texttt{BACKPHOTO\_THICK} & 24 \\

\texttt{WEIGHT\_TYPE} & NONE \\            
    
\texttt{WEIGHT\_IMAGE} & NONE \\
\hline
\end{tabular}
\caption{\textsc{SExtractor} configuration used throughout this paper.}
\label{tab:sextractor-config}
\end{table}

\section{Template spectra \& NYU Value-Added Galaxy Catalog}
\label{sec:nyu-vagc}

In the following, we elaborate on the template spectra used in the process of assigning spectral energy distribution to galaxies in \textsc{UFig} and on the corresponding model for the coefficients (see section \ref{subsubsec:gal-spectrum-model}). The templates are based on the Bruzual-Charlot stellar evolution synthesis models \cite{Bruzual2003}, they are the same ones as used by \texttt{kcorrect} \cite{Blanton2006} to estimate absolute magnitudes from observed fluxes and redshifts. To develop the model presented in \ref{subsubsec:gal-spectrum-model}, we used the New York University Value-Added Galaxy Catalog (NYU-VAGC, see \cite{Blanton2005}), a publicly available\footnote{\url{http://sdss.physics.nyu.edu/vagc/}} compilation of various surveys which were matched to SDSS. This highly complete catalog of local galaxies with redshifts mostly smaller than $z \sim 0.3$ contains five coefficients per galaxy which characterize the spectrum in terms of the same template spectra that we base our approach on.

To use this information in the process of building our model, we first split the catalog into two distinct populations according to the specific star formation rates (SSFRs) over the past 0.3$\,$Gyr, which are also given in the NYU-VAGC. Galaxies with $\text{SSFRs} > 0.007$ are considered as blue, galaxies with $\text{SSFRs} < 0.001$ are considered as red. Furthermore, we only keep galaxies which are included in the SDSS main galaxy spectroscopic sample, see \cite{Strauss2002}. This results in two samples of coefficients, which we fitted by Dirichlet distributions after further post-processing. We apply the following operations separately to the blue and to the red population:

\begin{itemize}

\item The means of the coefficients $c_i$ are divided out along each dimension.

\item Each sample is divided by its sum, s.t. all samples individually sum to 1.

\item We exclude all galaxies where any of the $c_i$ is within the lower 2\% of the range of values along the corresponding dimension. This yields a sub-sample which can be modeled much easier than the whole sample.

\item We re-weight the surviving sample along each dimension separately and then fit\footnote{Package used for fitting: \url{https://github.com/ericsuh/dirichlet}} it with a Dirichlet distribution. The weights are chosen such that the standard deviation of the five parameters of the Dirichlet distribution is minimized, which improves the quality of the fit.

\end{itemize}

This prescription provides us with two Dirichlet distributions of order five, whereby one model describes the blue and the other model the red population of galaxies. We do not use these best-fit results, instead, we vary the corresponding parameters during the ABC analysis (see section \ref{subsubsec:gal-spectrum-prior}). However, we additionally obtain the means that were divided out as well as the applied weights. These two are important, because the corresponding operations effectively change the colors of each galaxy in the fitted samples. Therefore, we store the means and weights such that they can be re-applied in order to preserve the shapes of the spectra. Hence, we effectively multiply each template spectrum by a fixed number such that the samples of coefficients can be fit by Dirichlet distributions. These numbers are different for blue and red galaxies. Thus, we effectively use different template spectra for the two different populations. We retain this prescription when sampling galaxies in \textsc{UFig}, i.e., we apply the same five numbers per population that were described above.

\section{Extinction map \& extinction law}
\label{sec:extinction}

Here, we briefly describe galactic extinction and the data we rely on to model this effect. Absorption and emission by the interstellar medium of the Milky Way cause galaxies to appear redder. In general, the interstellar medium absorbs light at UV wavelengths and re-emits at larger wavelengths, thus tilting the spectrum of galaxies towards the IR regime. Quantitatively, interstellar reddening can be accounted for by modifying the spectral energy distribution of a galaxy according to eq. \eqref{eq:extinction}. The wavelength-dependent extinction coefficient $A_\lambda$ varies with the line-of-sight and is generally larger for smaller wavelengths. To obtain $A_\lambda$ for the whole sky, the extinction map given by \cite{Schlegel1998} is used. It records the excess in $B-V$-color $E(B-V)$ due to reddening, which can be used to compute $A_\lambda$ via an extinction law. We use the extinction law originally proposed by \cite{Fitzpatrick1999} and recommended in combination with the extinction map from \cite{Schlegel1998} by \cite{Schlafly2010, Schlafly2011} in order to model interstellar reddening.

\end{document}